# RISMiCal: A software package to perform fast RISM/3D-RISM calculations

*Yutaka Maruyama,[1,2] and Norio Yoshida[3*]*

1 Data Science Center for Creative Design and Manufacturing, The Institute of Statistical Mathematics, 10-3 Midori-cho, Tachikawa, Tokyo 190-8562, Japan

2 Department of Physics, School of Science and Technology, Meiji University, 1-1-1 Higashi-Mita, Tama-ku, Kawasaki-shi, Kanagawa 214-8571, Japan

3 Graduate School of Informatics, Nagoya University, Chikusa, Nagoya 464-8601, Japan

**Corresponding Author**

*Norio Yoshida noriwo@nagoya-u.jp

**ABSTRACT**

Solvent plays an essential role in a variety of chemical, physical, and biological processes that occur in the solution phase. The reference interaction site model (RISM) and its three-dimensional extension (3D-RISM) serve as powerful computational tools for modeling solvation effects in chemical reactions, biological functions, and structure formations. We present the RISM integrated calculator (RISMiCal) program package, which is based on RISM and 3D-RISM theories with fast GPU code. RISMiCal has been developed as an integrated RISM/3D-RISM program that has interfaces with external programs such as Gaussian16, GAMESS, and Tinker. Fast 3D-RISM programs for single- and multi-GPU codes written in CUDA would enhance the availability of these hybrid methods because they require the performance of many computationally expensive 3D-RISM calculations. We expect that our package can be widely applied for chemical and biological processes in solvent. The RISMiCal package is available at https://rismical-dev.github.io.

**Introduction**

A liquid phase is the setting for various chemical and biological processes, where the solvent is one of the main actors, not a supporting player, determining the direction of these processes.[1-3] For example, in the molecular recognition process in solution, which is one of the most important fundamental biological processes, one must take into account not only the direct interaction between the host and guest molecules, but also the desolvation that accompanies the binding.[4,5] Prior to molecular recognition, the host and guest molecules are individually present in solution and interact with the solution molecules. In order for the guest molecule to bind to the binding site of the host molecule, the solution molecules present at the binding site must be expelled, and at the same time, the molecules in contact with the guest molecule must be excluded. Accompanying this desolvation is reconstruction of the interactions between solution molecules. In other words, molecular recognition is not a local process between host and guest molecules, but rather, a global process involving a change in the configuration of the entire system.

To describe such global changes in solution structure, the integral equation theory (IET) of liquids is the most suitable tool.[6-8] The molecular Ornstein–Zernike (MOZ) approach,[9-15] the reference interaction site model (RISM),[16-19] three-dimensional (3D)-RISM theories,[20,21] and various extensions[22-29] have been proposed as IETs for dealing with molecular solutions. In these theories, intermolecular correlation functions that account for many-body interactions between the components of solution can be obtained by using the pair interaction potentials between molecules as inputs. From the obtained correlation and distribution functions, various solvation thermodynamic quantities can be calculated, such as solvation free energy (SFE), solvation entropy, pressure, and partial molar volume.[7,8] The IETs have been used to study various chemical and biological processes in solution.

Although several types of IETs have been derived from treating intermolecular interactions in different ways, the 3D-RISM theory is one of the most successful because of its applicability and numerical robustness.[7,20,30,31] The 3D-RISM theory targets the solvation of infinitely diluted solute molecules and uses an interaction site model to obtain an approximate solvent distribution function under the external field created by the solute molecule. This formulation allows us to consider the solvation structure and thermodynamics of complicate molecular systems such as protein,[32-40] DNA,[41,42] telomere,[43] metal–organic frameworks,[44,45] polymers,[46-48] and so on. The representative example to show the power of the 3D-RISM theory is to find solvent molecules buried in proteins. Imai et al.[49] reported that the water distribution inside the hen-egg white lysozyme is in good agreement with experimental observations. Following these observations, the authors applied the 3D-RISM theory to mutated human lysozymes with different ion binding affinities and showed the applicability of the 3D-RISM theory to the prediction of the ion binding selectivity of proteins.[32] These studies led to subsequent molecular recognition studies based on the 3D-RISM theory. Examples of application are the analysis of ion and solvent permeation processes of channel proteins.[50-58] The free energy surfaces involved in molecular transport in the channel are quite complex, so the evaluation of the distribution function by the 3D-RISM is effective.

As these studies showed that the placement of solvent molecules could be predicted with high accuracy, methods were also developed to search for solvent molecules using the 3D-RISM. One such method is Placevent, developed by Sindhikara et al.[59] This method was simple enough to place the solvent molecules in the high peak positions of the solvent distribution. However, it accurately predicted the positions of water and ions and was therefore used to prepare the initial structure for molecular simulations. Since then, various improvements have been made and methods proposed to predict the positions and orientations of complex ligand molecules as well as ions and water.[60-63]

Another advantage of the 3D-RISM is that the solvation thermodynamic quantities can be calculated easily based on the distribution and correlation functions, such as SFE, solvation entropy, and partial molar volume. Because the 3D-RISM encompasses approximations related to the treatment of bridge functions and the solvent molecules orientation averaging by introducing interaction point models, the thermodynamic quantities obtained will include theoretical errors. Therefore, efforts have been made to improve the accuracy of the SFE estimates by adding corrections for molecular orientation and bridge functions.[62,64-67] The most successful correction method is the partial molar volume (PMV)-based correction method, and the first proposed method is called the universal collection (UC) method.[68] The UC method was developed based on the fact that the error in the non-electrostatic term of the SFE is proportional to the value of PMV calculated by the 3D-RISM. Several correction methods have been proposed following the UC method,[69,70] and recently, a method that eliminates fitting parameters by using pressure in addition to PMV, known as the pressure correction (PC) method, has been proposed.[71] Apart from this collection term, Sumi et al. have proposed a SFE functional for the 3D-RISM based on density functional theory (DFT).[72,73] With these correction methods, the SFE obtained by the 3D-RISM reached the same level of accuracy as that obtained by the free energy perturbation method.

An important extension of the 3D-RISM is the development of hybrid methods with other computational science methods. As a hybrid method with electronic structure theory, the RISM-self-consistent field (SCF) method was first developed by Ten-no et al.[74,75] In this method, the distribution function of the solvent and electron density distribution of the solute are determined to be consistent with each other. Subsequently, hybrid methods of the electronic structure theory with the 3D-RISM and MOZ, referred as the 3D-RISM-SCF[21,76] and MOZ-SCF,[77,78] respectively, have been developed. Because this formalism is quite general, it has been proposed to be combined with not only the Hartree–Fock (HF) method, but also post-HF methods represented by multi-configurational (MC) SCF and DFT, and has been applied to not only molecular ground, but also excited states. Hybrid methods with the quantum mechanics/molecular mechanics (QM/MM) and fragment molecular orbital (FMO) methods were also proposed to handle macromolecules.[42,79] Various applications of these hybrid methods have been made to chemical reactions in solution.

The hybrid method with molecular simulation allows us to take structural fluctuations and changes of solute molecules into account. As a combination of simulation and IET, a method combining Monte Carlo and RISM methods has been proposed by Kinoshita et al.[80,81] This method uses the SFE calculated by the 3D-RISM and the structural energy of the solute calculated by MM force field to determine the Metropolis for the structural change of the solute. Following the combination with the stochastic MC method, a combination with the deterministic molecular dynamics (MD) method was proposed by Miyata et al.,[82] which is referred to as MD/3D-RISM method. In the MD/3D-RISM method, the computational cost of the 3D-RISM is quite large for one MD step, so the number of 3D-RISM calculations is reduced by using the multi-time-step method. Because the solvent-derived forces obtained from the 3D-RISM calculation are approximated by extrapolation, there is a problem that the larger the width of the multi-time step, the larger the error becomes.[83-85] Therefore, a large time-step width may cause problems, such as the inability to obtain an accurate solute structure ensemble corresponding to the 3D-RISM-based Hamiltonian. Recently, methods have been developed to reproduce ensembles for accurate 3D-RISM Hamiltonians based on hybrid Monte Carlo (HMC) methods, which is referred to as the HMC/3D-RISM method.[86] Both the hybrid method with the electronic structure theory and molecular simulations require iterative 3D-RISM calculations many times, so a fast 3D-RISM program is desirable.

A number of RISM/3D-RISM programs are currently available. The RISM/3D-RISM program in one of the most representative programs, AmberTools, is useful for analyzing the biomolecular systems solvation in combination with the Amber MD simulation package developed by Luchko et al.[83] EPISOL, developed by Cao et al.,[87] is another program for the application of 3D-RISM for a biomolecular system. For a supercomputer system, *the RISM for HPC* was developed by Wilson, Kobryn, and Sergey and applied to large biomolecular systems.[88,89] The RISM/3D-RISM program has also been implemented in quantum chemistry program packages such as Amsterdam Density Functional,[90] SALMON,[91] Quantum ESPRESSO[92], and GAMESS.[93] These packages have been widely used in solvation studies of various molecular systems. However, with the developments in computational molecular science, solvation calculations for more complex systems are required, and as mentioned above, the development of even faster 3D-RISM programs is needed.

In this paper, we introduce the open-source program code of RISM/3D-RISM, referred to as RISM integrated calculator (RISMiCal). The RISMiCal code was a proprietary code that has been developed by the authors and includes the various RISM, 3D-RISM, and MOZ codes and their extended forms described above.[94] The programs to be released as open source include the basic code for the RISM and 3D-RISM and the fast code for single- and multi-GPU.[95,96] The authors have also developed programs for massively parallel supercomputers, such as K computer and supercomputer Fugaku.[97] The bottleneck in parallelizing 3D-RISM calculations is the 3D Fast Fourier Transform (3D-FFT). In parallel 3D-FFT across computer nodes, communication associated with axis exchange is the rate-limiting factor. To reduce the amount of communication involved in axis exchange, we proposed a pencil parallel 3D-FFT. This method has achieved high parallelization efficiency even in massively parallel computers such as the K computer. The multi-GPU code included in RISMiCal also employs this technique, which achieved high parallelization efficiency on multi-GPU, thereby enabling the computation of large systems such as protein complexes.

The remainder of this paper is organized as follows. First, the RISM and 3D-RISM theories are outlined. Then, the hybrid method MM/3D-RISM and 3D-RISM-SCF formalisms are presented. Next, numerical procedures and methods for accelerating convergence are described. Parallelization methods using two-dimensional (2D) distribution 3D-FFT will also be discussed. The features of this package, such as computational scheme and usage, are explained.  As an example of testing this program, results are shown for the hydration structure of the coronavirus spike protein and human angiotensin-converting enzyme 2 (ACE2) complex. Finally, the results of a benchmark using this system as an example are shown.

**Theory**

In this section, the basics of the RISM and 3D-RISM theories and those extension methods are reviewed. The RISM and 3D-RISM theories are derived from the MOZ equation by taking the orientational averaging centered on atoms. The MOZ equation for multicomponent molecular solution is given by

$$h_{IJ}(1,2) = c_{IJ}(1,2) + \sum_K \rho_K \int c_{IJ}(1,3) h_{KJ}(3,2) d(3) \tag{1}$$

where $c_{IJ}(1,2)$ and $h_{IJ}(1,2)$ are the direct and total correlation functions between the components $I$ and $J$ of solution, and the numbers 1, 2, and 3 denote the orientation and coordinates of molecules 1, 2 and 3, respectively. $d(3)$ denotes the integral with respect to the position and orientation of molecule 3. $\rho_K$ is the number density of component $K$ and the summation over $K$ is running all the components in solution.

***RISM theory*** In the RISM theory, the correlation functions are expressed as a function between interaction sites, referred to as site–site correlation functions.[16-18] The site–site total correlation functions are defined as

$$h_{\alpha\gamma}(r) = \frac{1}{\Omega^2} \int h_{IJ}(1,2) \delta(|\mathbf{r}_1 + \mathbf{l}_1^\alpha|) \delta(|\mathbf{r}_2 + \mathbf{l}_2^\gamma| - r) d(1) d(2) \tag{2}$$

where $\alpha$ and $\gamma$ indicate interaction sites on atoms $I$ and $J$, respectively. $\Omega$ means normalized constant of orientational averaging, and $\mathbf{l}_1^\alpha$ and $\mathbf{l}_2^\gamma$ are vectors connecting the molecular center and interaction site. $d(1)$ and $d(2)$ denote the integrals with respect to the orientation and positions of molecules 1 and 2, respectively. By applying this transformation and approximation of the direct correlation function

$$C_{IJ}(1,2) \approx \sum_{\alpha\gamma} c_{\alpha\gamma}(r), \tag{3}$$

to the MOZ equation, (1), the RISM equation can be obtained as

$$h_{\alpha\gamma}(r) = \omega_{\alpha\alpha'} * c_{\alpha'\gamma'} * (\omega_{\gamma'\gamma} + \rho_{\gamma'}h_{\gamma'\gamma}) \tag{4}$$

where $\omega_{\alpha\alpha'}$ denotes an intramolecular correlation function. The symbol * indicates that a convolution integral and the summation should be taken for repeated indices. To close this equation, another equation representing the relation of the correlation function that appeared in Eq. (4) is required. To date, many closure equations have been proposed, such as hyper-netted chain (HNC), Percus–Yevick, Kovalenko–Hirata (KH), and so on. To the best of our knowledge, the KH closure is most widely used for the RISM study because of its robustness regarding the computational convergence.[21,98] The HNC closure relation is given by

$$g_{\alpha\gamma}(r) = \exp[-\beta u_{\alpha\gamma}(r) + h_{\alpha\gamma}(r) - c_{\alpha\gamma}(r)] \tag{5}$$

and the KH closure relation is given by

$$g_{\alpha\gamma}(r) = \begin{cases} \exp[d_{\alpha\gamma}(r)] & \text{for } d_{\alpha\gamma}(r) < 1 \\ d_{\alpha\gamma}(r) + 1 & \text{for } d_{\alpha\gamma}(r) \geq 1 \end{cases}, \tag{6}$$

where

$$d_{\alpha\gamma}(r) = -\beta u_{\alpha\gamma}(r) + h_{\alpha\gamma}(r) - c_{\alpha\gamma}(r). \tag{7}$$

Here, $g_{\alpha\gamma}(r) = h_{\alpha\gamma}(r) + 1$ is a radial distribution function between interaction sites $\alpha$ and $\gamma$, and $u_{\alpha\gamma}$ is a pair interaction potential between sites $\alpha$ and $\gamma$. $\beta = 1/k_B T$ is inverse temperature. By solving the RISM and closure equations under a given $u_{\alpha\gamma}$ iteratively, one can obtain the distribution and direct correlation functions.

When focusing on a single molecule immerged in solution at infinite dilution, Eq. (1) can be simplified as follows:

$$h_{\alpha\gamma}(r) = \omega_{\alpha\alpha'} * c_{\alpha'\gamma'} * X_{\gamma'\gamma} \tag{8}$$

where $\alpha$ and $\alpha'$ belong to the focusing molecule referred to "solute" molecule, and $\gamma$ and $\gamma'$ belong to all other species in solution referred to "solvent" molecules. $X_{\gamma'\gamma} = \omega_{\gamma'\gamma} + \rho_{\gamma'}h_{\gamma'\gamma}$ is the solvent susceptibility function, which can be obtained by solving the RISM equation for a pure solvent system.

By solving these equations, various thermodynamic quantities such as SFE can be obtained from correlation functions. The analytical expression of the SFE for RISM/HNC is derived by Singer and Chandler[99] as

$$\Delta\mu = \frac{4\pi}{\beta} \sum_{\alpha\gamma} \int \left[ -c_{\alpha\gamma}(r) - \frac{1}{2}h_{\alpha\gamma}(r)c_{\alpha\gamma}(r) + \frac{1}{2}h_{\alpha\gamma}^2(r) \right] r^2 dr \tag{9}$$

and similarly, that for RISM/KH[21] is given by

$$\Delta\mu = \frac{4\pi}{\beta} \sum_{\alpha\gamma} \int \left[ -c_{\alpha\gamma}(r) - \frac{1}{2}h_{\alpha\gamma}(r)c_{\alpha\gamma}(r) + \frac{1}{2}h_{\alpha\gamma}^2(r)\Theta\left(-h_{\alpha\gamma}(r)\right) \right] r^2 dr \tag{10}$$

where $\Theta$ denotes the Heaviside step function.

***3D-RISM theory*** The 3D-RISM deals with 3D correlation functions of solvents in the fields created by solute molecules.[20,30,31] In the case of the 3D-RISM, the 3D total correlation function is defined as

$$h_\gamma(\boldsymbol{r}) = \frac{1}{\Omega}\int h_{IJ}(1,2)\delta(|\boldsymbol{r}_2 + \boldsymbol{l}_2^\gamma - \boldsymbol{r}|)d(2) \tag{11}$$

Unlike Eq. (2), an orientational average is taken only for the orientation of the solvent molecule. For the direct correlation function, the following approximation is introduced:

$$C_{IJ}(1,2) \approx \sum_\gamma c_\gamma(\boldsymbol{r}). \tag{12}$$

Using these equations, the following 3D-RISM equations can be derived:

$$h_\gamma(\boldsymbol{r}) = \sum_{\gamma'} c_{\gamma'} * X_{\gamma'\gamma}, \tag{13}$$

where $X_{\gamma'\gamma}$ is the solvent susceptibility function for pure solvent, which must be obtained by solving the RISM equation for a pure solvent system prior to the 3D-RISM calculation. The closure relation for the 3D-RISM equation has also been proposed. For example, the HNC closure is given by

$$g_\gamma(\boldsymbol{r}) = \exp[-\beta u_\gamma(\boldsymbol{r}) + h_\gamma(\boldsymbol{r}) - c_\gamma(\boldsymbol{r})], \tag{14}$$

and the KH closure by

$$g_\gamma(\boldsymbol{r}) = \begin{cases} \exp\{d_\gamma(\boldsymbol{r})\} & d_\gamma(\boldsymbol{r}) > 0 \\ d_\gamma(\boldsymbol{r}) + 1 & d_\gamma(\boldsymbol{r}) \leq 0 \end{cases}, \tag{15}$$

$$d_\gamma(\boldsymbol{r}) = -\beta u_\gamma(\boldsymbol{r}) + h_\gamma(\boldsymbol{r}) - c_\gamma(\boldsymbol{r}), \tag{16}$$

where $u_\gamma(r)$ is an interaction potential due to the solute molecule acting on the solvent site $\gamma$ at position $r$. The SFE of the 3D-RISM/HNC is given by

$$\Delta\mu = \frac{1}{\beta}\sum_\gamma \rho_\gamma \int \left[-c_\gamma(r) - \frac{1}{2}h_\gamma(r)c_\gamma(r) + \frac{1}{2}\{h_\gamma(r)\}^2\right] dr \qquad (17)$$

and

$$\Delta\mu = \frac{1}{\beta}\sum_\gamma \rho_\gamma \int \left[-c_\gamma(r) - \frac{1}{2}h_\gamma(r)c_\gamma(r) \right.$$
$$\left. + \frac{1}{2}\{h_\gamma(r)\}^2 \Theta\{-h_\gamma(r)\}\right] dr \qquad (18)$$

for the 3D-RISM/KH case.[21] In addition to SFE, one can obtain various thermodynamic properties such as PMV, pressure, internal energy, and solvation entropy. The PMV is obtained by

$$\bar{V} = k_B T \chi_T \left(1 - \sum_\gamma \rho_\gamma \int c_\gamma(r) dr\right) \qquad (19)$$

where $\chi_T$ is the isothermal compressibility that can be obtained by the RISM calculation for a pure solvent system.[100] The pressure of the system can be evaluated several ways based on the RISM/3D-RISM results. One of these ways is defined by the work, namely free energy change, required to exclude the solvent from the macroscopic volume, starting from a uniform density solvent.[67,71] In this way, the bulk pressure is given by

$$P = \sum_\Gamma \frac{\rho_\Gamma}{2\beta}(n_\Gamma + 1) - \sum_{\gamma\lambda} \frac{2\pi\rho_\gamma\rho_\lambda}{\beta} \int c_{\gamma\lambda}(r) r^2 dr \qquad (20)$$

where $c_{\gamma\lambda}$ is the direct correlation function for the bulk solvent. The summation of the first term on the right-hand side with respect to $\Gamma$ is running over the solvent species, and $\rho_\Gamma$ and $n_\Gamma$ are the number density and number of solvent sites of species Γ, respectively. From the PMV and pressure, one can obtain the PC term for the SFE

$$\Delta\mu_{\text{PC}} = -P\bar{V} \tag{21}$$

which is well known to drastically improve the accuracy of the SFE value. There quantities can be obtained by using the RISMiCal package.

**Hybrid implementation**

Hybrid methods have been proposed that combine the 3D-RISM theory with molecular simulation and quantum chemical electronic structure theory. The RISMiCal package has an interface to those software programs.

*MM/3D-RISM* The 3D-RISM theory is solved under a given solute structure and potential. In the MM context, the solute–solvent interaction potential is given by

$$u_\gamma(\mathbf{r}) = \sum_\alpha \left[ \frac{q_\alpha q_\gamma}{|\mathbf{r} - \mathbf{R}_\alpha|} + 4\varepsilon_{\alpha\gamma} \left\{ \left( \frac{\sigma_{\alpha\gamma}}{|\mathbf{r} - \mathbf{R}_\alpha|} \right)^{12} - \left( \frac{\sigma_{\alpha\gamma}}{|\mathbf{r} - \mathbf{R}_\alpha|} \right)^6 \right\} \right] \tag{22}$$

where $q_\alpha$ and $\mathbf{R}_\alpha$ denote the point charge and position of solute atom $\alpha$. $\varepsilon_{\alpha\gamma}$ and $\sigma_{\alpha\gamma}$ are the Lennard–Jones (LJ) parameters with usual meanings. The total free energy of the system for a given solute structure is defined as

$$G_{\text{MM}}(\{\mathbf{R}_\alpha\}) = E_{\text{MM}}(\{\mathbf{R}_\alpha\}) + \Delta\mu(\{\mathbf{R}_\alpha\}) \tag{23}$$

where $E_{\text{MM}}$ is solute structural energy.[82] An analytical expression of the first derivative of the free energy respect to the coordinate of solute atom is derived as[90,101]

$$\frac{\partial G_{\text{MM}}\{\boldsymbol{R}_\alpha\}}{\partial R_\alpha} = \frac{\partial E_{\text{MM}}\{\boldsymbol{R}_\alpha\}}{\partial R_\alpha} + \sum_\gamma \rho_\gamma \int \frac{\partial u_\gamma(\boldsymbol{r})}{\partial R_\alpha} g_\gamma(\boldsymbol{r}) d\boldsymbol{r}. \tag{21}$$

This expression is common for both HNC and KH. This formula allows us to perform the geometry optimization or MD simulation under the solvent environment described by the 3D-RISM theory.

***3D-RISM-SCF*** A hybrid of the 3D-RISM and the quantum mechanics electronic structure theory is referred as 3D-RISM-SCF theory because the electronic structure and solvent distribution are determined simultaneously to be consistent.[21,76] Similarly, hybrids of RISM and MOZ with quantum mechanics theory are called RISM-SCF[74,75,102] and MOZ-SCF,[77,78] respectively. Here, we briefly explain a basis of 3D-RISM-SCF.

In 3D-RISM-SCF, similar to Eq. (13), the free energy of the system is defined by

$$G_{\text{SCF}}(\{\boldsymbol{R}_\alpha\}) = E_{\text{QM}}(\{\boldsymbol{R}_\alpha\}) + \Delta\mu(\{\boldsymbol{R}_\alpha\}) \tag{24}$$

where $E_{\text{QM}}(\{\boldsymbol{R}_\alpha\})$ is the solute energy with a given solute structure defined by the set of atomic coordinates $\{\boldsymbol{R}_\alpha\}$, which is given by

$$E_{\text{QM}}(\{\boldsymbol{R}_\alpha\}) = \langle \Psi | \hat{H}_0 | \Psi \rangle + E_{\text{nuclei}}, \tag{25}$$

where $\Psi$ and $\hat{H}_0$ are solute electron wave function in solution and the Hamiltonian in an isolated system. $E_{\text{nuclei}}$ is the internuclear electrostatic repulsion energy. The solute wave function $\Psi$ is obtained by solving the following Schrödinger's equation

$$(\hat{H}_0 + \hat{V}_{\text{solv}})|\Psi\rangle = E_{\text{solv}}|\Psi\rangle, \tag{26}$$

where $\hat{V}_{\text{solv}}$ is an electrostatic potential acting on a solute electron because of the solvent distribution

$$\hat{V}_{\text{solv}}(\bm{r}_e; \{\bm{R}_\alpha\}) = -\sum_\gamma \rho_\gamma \int \left(\frac{q_\gamma}{|\bm{r}_e - \bm{r}|}\right) g_\gamma(\bm{r}) d\bm{r}. \tag{27}$$

The solvent distribution function $g_\gamma(\bm{r})$ is determined by solving the 3D-RISM under the potential resulting from the solute electron distribution and nuclear charge:

$$u_\gamma(\bm{r}) = \sum_\alpha \left[\frac{Z_\alpha q_\gamma}{|\bm{r} - \bm{R}_\alpha|} - \int \frac{|\Psi(\bm{r}_e)|^2 q_\gamma}{|\bm{r}_e - \bm{r}|} d\bm{r}_e + 4\varepsilon_{\alpha\gamma}\left\{\left(\frac{\sigma_{\alpha\gamma}}{|\bm{r} - \bm{R}_\alpha|}\right)^{12} - \left(\frac{\sigma_{\alpha\gamma}}{|\bm{r} - \bm{R}_\alpha|}\right)^{6}\right\}\right], \tag{28}$$

where $\bm{r}_e$ denotes the electron coordinates. It is noted that the LJ potential is employed to reproduce the exchange repulsion and van der Walls interaction between solute and solvent molecules.

**Numerical procedure**

Because the RISM and 3D-RISM are nonlinear coupled equations, they should be solved numerically in an iterative manner. Two efficient methods are known for solving the RISM/3D-RISM: iterative calculation using $c(r)$ or $t(r)(= h(r) - c(r))$. We have adopted the method using $t(r)$ because the convergence is more stable.[7,12,103,104] The 3D-RISM solution method is described here; for the 1D-RISM solution method, see the description by Kinoshita.[7]

To handle the long-range Coulombic potentials, the following long-range parts of the functions are defined:

$$f_\gamma(\bm{r}) = \sum_\alpha \frac{\beta q_\alpha q_\gamma}{|\bm{r} - \bm{R}_\alpha|}(1 - \exp(-|\bm{r} - \bm{R}_\alpha|)), \tag{29}$$

and

$$\tilde{f}_\gamma(\bm{k}) = \sum_\alpha \frac{4\pi\beta q_\alpha q_\gamma}{k^2(k^2 + 1)} e^{-i\bm{k}\cdot\bm{R}_\alpha}, \tag{30}$$

where $R_\alpha$ is the position of solute site $\alpha$, $q_\alpha$ and $q_\gamma$ are the partial charges on solute site $\alpha$ and solvent site $\gamma$, respectively, and "~" represents Fourier transforms. To avoid the singularity of $\tilde{f}_\gamma(k)$ at $k = 0$, the grids in $k$-space are placed at half-integer points, namely $k_i = \Delta k(i + \frac{1}{2})$, whereas those in $r$-space are placed at integer points. This allows the divergent terms, $f_\gamma(r)$ and $\tilde{f}_\gamma(k)$, to be treated analytically.

The procedure for solving the 3D-RISM theory is as follows:

1. Choose initial guesses of $t_\gamma(r)$, $t_\gamma(r) = f_\gamma(r)$.

2. From $t_\gamma(r)$, calculate $g_\gamma(r)$ using the HNC/KH closure equation. Then, calculate the short-range parts of the $c_\gamma(r)$, $c_\gamma^{sr}(r)$ from the following:

$$c_\gamma^{sr}(r) = g(r)_\gamma - 1 - t_\gamma(r) + f_\gamma(r). \tag{31}$$

3. Obtain the Fourier transform $\tilde{c}_\gamma^{sr}(k)$, then calculate $\tilde{c}_\gamma(k)$,

$$\tilde{c}_\gamma(r) = \tilde{c}_\gamma^{sr}(k) - \tilde{f}_\gamma(k). \tag{32}$$

4. Calculate $\tilde{h}_\gamma(k)$ using the 3D-RISM equation.

5. Obtain $h_\gamma(r)$ via the backward Fourier transformation. Then calculate $t'_\gamma(r)$,

$$t'_\gamma(r) = t_\gamma(r) + h_\gamma(r) + 1 - g_\gamma(r). \tag{33}$$

6. Determine the new input values $t_\gamma(r)$ from $t_\gamma(r)$ and $t'_\gamma(r)$ using an acceleration method, e.g., the modified procedure of the direct inversion in the iterative subspace method, or the modified Anderson method.

7. Repeat steps 2-6 until the input and output values of the iteration variables become identical within convergence tolerance.

**Convergence acceleration**

Due to the high computational cost of the 3D-RISM calculations, several convergence acceleration methods have been proposed to reduce the number of convergence times: the Picard iteration with dynamic relaxation (the dynamic relaxation method),[20] the modified procedure of the direct inversion in the iterative subspace (MDIIS),[104] the multigrid algorithm,[105] the modified Anderson method,[95] and the static mixing approach.[87] Among these, the MDIIS is the most widely used. In our package, the MDIIS is used for the FORTRAN code and the modified Anderson method for the C++/CUDA code for GPUs.

*Modified Anderson*

We developed a modified Anderson method to perform the 3D-RISM calculations on GPUs. A 3D-RISM calculation for water solvents on a $256^3$ grid with double precision and an MDIIS subspace set to 10 requires a total of 9 GB of memory. On the other hand, GPUs in 2010 had a maximum of only 4 GB of memory, but had a wider memory bandwidth than CPUs. Various convergence acceleration methods were investigated to perform the 3D-RISM calculations without data transfer between CPUs and GPUs, resulting in the modified Anderson method. The convergence is comparable to the modified Anderson method with two histories and MDIIS with 10 subspaces.[95]

The Anderson method is a type of mixing method. Simple mixing methods do not vary the same mixing rate, whereas the Anderson method varies the mixing rate dynamically. Here, the modified Anderson method with *two* histories is presented. The distribution determined by the *n*th iterative calculation is denoted as $Y^{(n)}$, and the distribution updated by the Anderson method is denoted as $X^{(n)}$. In the RISM/3D-RISM solution scheme, the indirect correlation function, $t(r) = h(r) - c(r)$, is employed as $X(r)$. Define

$$u^{(n-1)}(r) = X^{(n-1)}(r)$$

$$+ \sum_{i=1}^{2} \left(\theta_i^{(n-1)} + s_i\right)\left(X^{(n-i-1)}(r) - X^{(n-1)}(r)\right), \tag{34}$$

$$v^{(n)}(r) = Y^{(n)}(r)$$

$$+ \sum_{i=1}^{2} \left(\theta_i^{(n-1)} + s_i\right)\left(Y^{(n-i)}(r) - Y^{(n)}(r)\right), \tag{35}$$

where $\theta_i^{(n)}$ is dynamical free parameter and $s_i$ is a tuning parameter not found in the original Anderson method. This tuning parameter rebalances the contributions of previous distribution functions and adapts the Anderson method to accelerate the convergence of the 3D-RISM calculations.

By solving the simultaneous equations for $\theta_i^{(n)}$,

$$\sum_{j=1}^{2} \left(R^{(n-1)}(r) - R^{(n-j-1)}(r), R^{(n-1)}(r) - R^{(n-i-1)}(r)\right)\theta_i^{(n-1)}$$

$$= \left(R^{(n-1)}(r), R^{(n-1)}(r) - R^{(n-j-1)}(r)\right) \tag{36}$$

for $i = 1, 2$. Here,

$$R^{(n-1)}(r) = Y^{(n)}(r) - X^{(n-1)}(r), \tag{37}$$

and an inner product of two vectors defined by

$$(a(r), b(r)) = \sum_r a(r)b(r)w \tag{38}$$

with weight factor, $w$. We choose the inverse of the number density, $1/\rho^v$, as the weight factor. Then, we can get the new $X^{(n)}(r)$ by

$$X^{(n)}(r) = u^{(n-1)}(r) + b\left(v^{(n)}(r) - u^{(n-1)}(r)\right), \tag{39}$$

where $b$ is a static mixing parameter. The optimal value of $b$ is empirically determined for values in the range 0 to 1. In the case of two histories, the Picard method is used for the first two steps and switched to the modified Anderson method from the third step.

*Modified DIIS*

According to the above symbols, MDIIS is expressed by

$$X^n(r) = Y^n(r) + b \sum_{i=1}^{M} c_i \left(X^{(n-i)}(r) - Y^{(n-i)}(r)\right), \tag{40}$$

where $c_i$ are coefficients decided by the Lagrange multiplier technique and $b$ is a damping parameter. If the mixing parameter $b$ is set to 1, the above formula becomes the original DIIS. A new trial function, $X^{(n)}(r)$, consists of $Y^{(n)}(r)$ and a linear combination of previous error vectors. The MDIIS method requires a large memory space to store 10 error vectors from previous steps for the 3D-RISM calculation.

In the original MDIIS, the damping parameter b was set to a static value. However, when this parameter was changed dynamically to $b = -0.2\log(R)$, where $R$ is the root-mean-square deviation of the residuals, the 3D-RISM calculation is stable and converges quickly.[97]

As of 2023, NVIDIA's top-of-the-range GPU, the H100, had 80 GB, which is sufficient to perform the 3D-RISM calculations using MDIIS for the convergence acceleration method. On the other hand, the RTX series, which is often used for MD calculations, has a minimum of 8 GB, which is not enough to run MDIIS.

**2D decomposition 3D-FFT**

Next, we describe the procedure for the 2D decomposition parallel 3D-FFT (pencil 3D-FFT) for $N (= P \times Q)$ nodes. The array is distributed to P (y-axis) × Q (z-axis) nodes on the calculation cell (Fig. 1 upper). The forward procedure for the normal pencil 3D-FFT is as follows:

1. Perform the 1D-FFT in the x-axis. (Fig. 1 upper)

2. Perform the collective communication for the Q-parallel set of P nodes.

3. Perform the 1D-FFT in the y-axis. (Fig. 1 middle)

4. Perform the collective communication for the P-parallel set of Q nodes.

5. Perform the 1D-FFT in the z-axis. (Fig. 1 bottom)

6. Perform the collective communication for the P-parallel set of Q nodes.

7. Perform the collective communication for the Q-parallel set of P nodes.

The distributed arrays of the real and Fourier spaces are the same after procedure 7. Here, procedures 6 and 7 are performed only to regulate the array. We can thus cancel these procedures to reduce the collective communication in the 3D-FFT. After procedure 5, the calculation cell is distributed along the x- and y-axes (Fig. 1 bottom). Then, the backward procedure in this case is as follows:

1. Perform the 1D-FFT in the z-axis.

2. Perform the collective communication for the P-parallel set of Q nodes.

3. Perform the 1D-FFT in the y-axis.

4. Perform the collective communication for the Q-parallel set of P nodes.

5. Perform the 1D-FFT in the x-axis.

We adapted the 3D-RISM program for multi-GPU to this distribution pattern in Fourier space. We chose the values of P and Q according to the following rule to balance the communication: $P = Q = 2^n$ when $N = 2^{2n}$, or $P = 2^n$ and $N = 2^{n+1}$ when $N = 2^{2n+1}$.

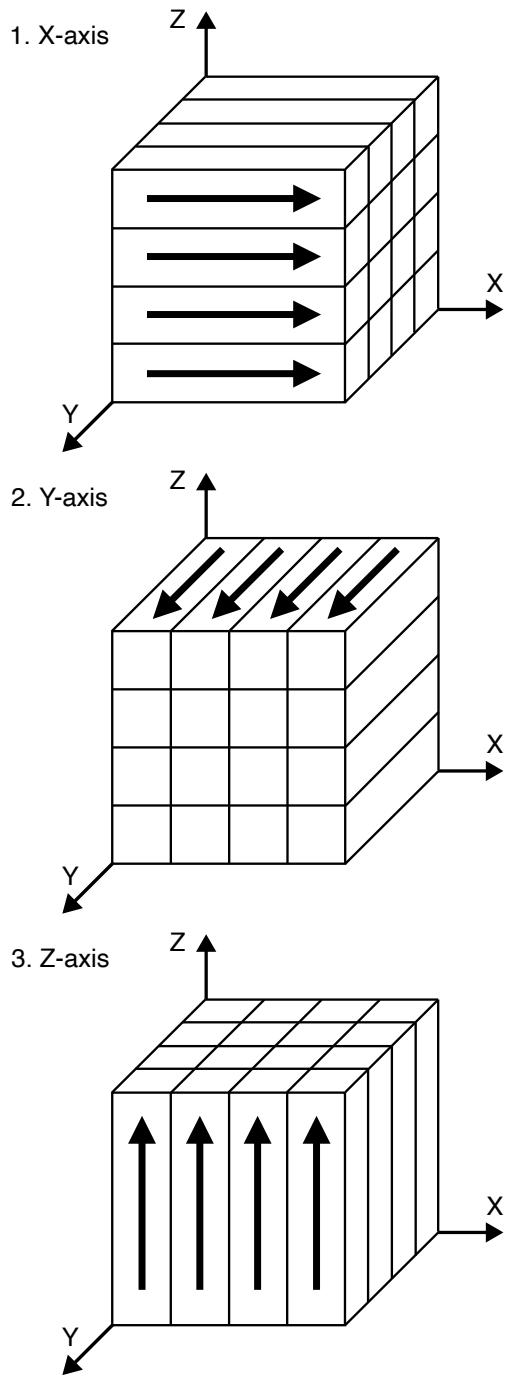

Figure 1. The two-dimensional decomposition pattern of the three-dimensional calculation cell in pencil 3D-FFT.

**Package Features**

In this section, the features of the RISMiCal package are explained. The RISMiCal package consists of three sets of programs: the main body of the RISMiCal code, a single GPU, and a multi-GPU version of the 3D-RISM engine. The main body of the RISMiCal code consists of an input/output user interface, the 1D-RISM for solvent–solvent and solute–solvent systems, and the 3D-RISM for solute–solvent systems. Scheme 1 presents an outline of the flowchart. The inputs and command line option should be read first. The calculations were performed according to the system type specified in the command line option. Usually, for solvent–solvent systems, the 1D-RISM is performed first, to calculate the solvent susceptibility, $\chi_{vv}$. This calculation does not need to be repeated if the same solvent is used. Next, the obtained solvent susceptibility is used to calculate the solute–solvent system. Although all the functions can be performed with this main package alone, faster GPU 3D-RISM engines are available if the user's system is equipped with GPUs. There are two types of GPU codes, one for a single GPU and one for multiple GPUs, and the choice is made according to the size of the system to be computed. If the number of grids exceeds 512 per side, the multi-GPU version may be selected to satisfy the memory requirements given the current on-board memory capacity of the GPUs. The GPU 3D-RISM engines require roughly 2.3 GB memory of GPUs per $256^3$ grids with one solvent site, and the memory requirement is almost proportional to the number of grids and solvent sites.

The package includes several example calculations, which users can learn how to perform. Detailed specifications of the inputs can be found in the Supporting Information or in the manual available on GitHub.

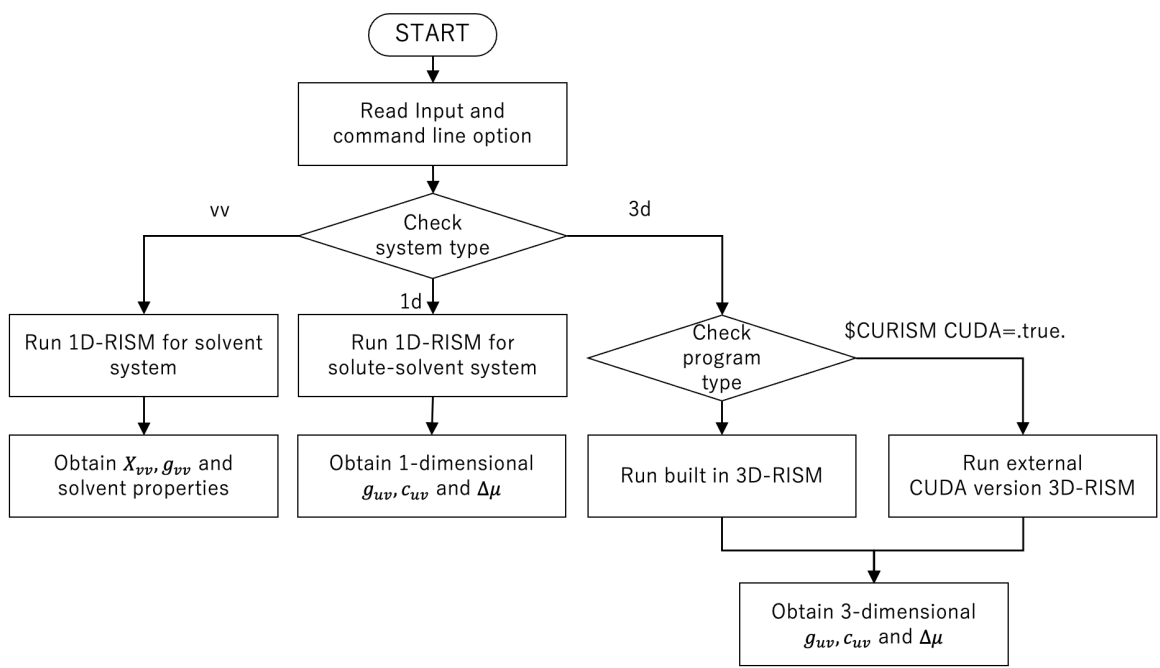

Scheme 1: Flowchart of the RISMiCal package.

**Example**

To demonstrate the performance of the RISMiCal package, the receptor-binding domain (RBD) of the severe acute respiratory syndrome coronavirus 2 spike protein complex with ACE2 (total 12877 atoms) in 0.2 M NaCl aqueous solution ($512^3$ grids, $256^3$ Å$^3$ box) is used as an example.[106] In Figure 2, the structure of the RBD–ACE2 complex generated by the MD simulation taken from our previous work[106] is depicted. This is one of the 8900 structures used in our previous study of the binding between RBD and ACE2, which was originally taken from the DESRES-ANTON trajectory dataset [10857295,10895671], available on the D. E. Shaw research website.[107] For the 3D-RISM calculation, we employed the AMBER ff14SB, GLYCAM 06j, and TIP3P force fields for proteins, glycans, and water, respectively.[108-110] The parameter assignment is performed by using the tLeap program in Amber 20, and the generated parameter files in AMBER format are converted into RISMiCal format using the amb2rsm tool in the RISMiCal. The parameters for the modified Anderson method are as follows: $s_1 = 0.2, s_2 = 0.0$, and $b = 0.6$.

The 3D distributions of the solvent species around the target RBD–ACE2 complex structure obtained by the 3D-RISM theory are demonstrated in Figure 3. It displays isosurfaces using two different thresholds, namely $g_\gamma(r) = 2.0$ and 4.0. The $g_O(r) = 2.0$ surface of oxygen in water appears to cover the entire protein. By contrast, the surfaces at $g_O(r) = 4.0$ are scattered, suggesting that they are oriented around specific amino acid atoms. Unlike the case of oxygen, the distribution of ions does not cover the entire protein, indicating strong coordination with specific residues. Namely, negatively charged residues are located in areas where the distribution of Na$^+$ is found, and positively charged residues are located in areas where the distribution of Cl$^-$ is found. From this distribution function, the positions of explicit solvent molecules can be predicted. The explicit solvent molecule positions calculated using one such method, the Placevent algorithm,[59] are demonstrated in Figure 4. In the ACE2-RBD contact region, it can be seen that the oxygen atoms in water make hydrogen bond to the amino group of lysine and the hydrogen of the hydroxyl group of serine. Coordination of the Na$^+$ ions to the carboxyl group of aspartic acid was also observed. On the other hand, Cl$^-$ ions did not appear on the contact region.

The ACE2-RBD complex was used as an example to demonstrate the solvent distribution obtained by the 3D-RISM. This system will be used throughout the following analysis of computational efficiency.

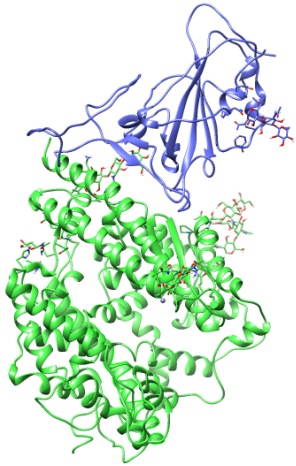

Figure 2. Structure of the RBD–ACE2 complex taken from a previous study.[106] RBD and ACE2 are depicted in blue and green, respectively.

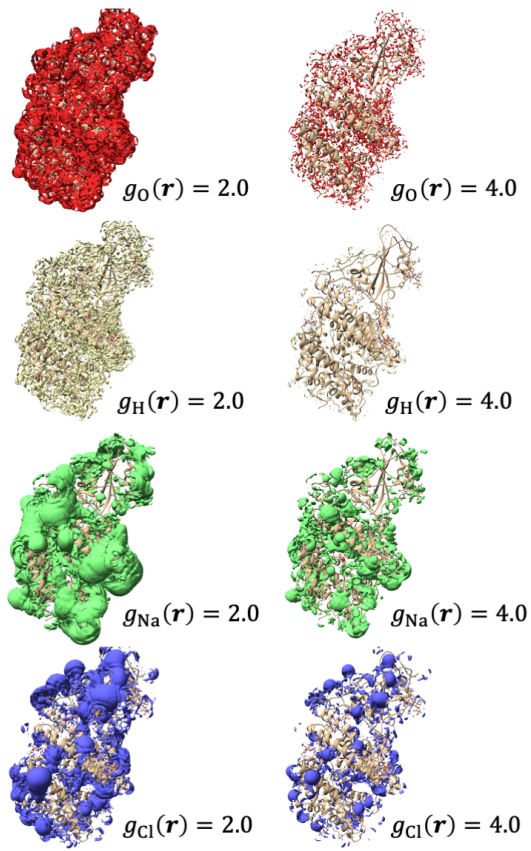

Figure 3. Iso-surface plots of solvent distributions. Oxygen and hydrogen in water, Na$^+$, and Cl$^-$ ions are plotted in red, white, green, and blue, respectively. The values of the iso-surface plots are given in the figure.

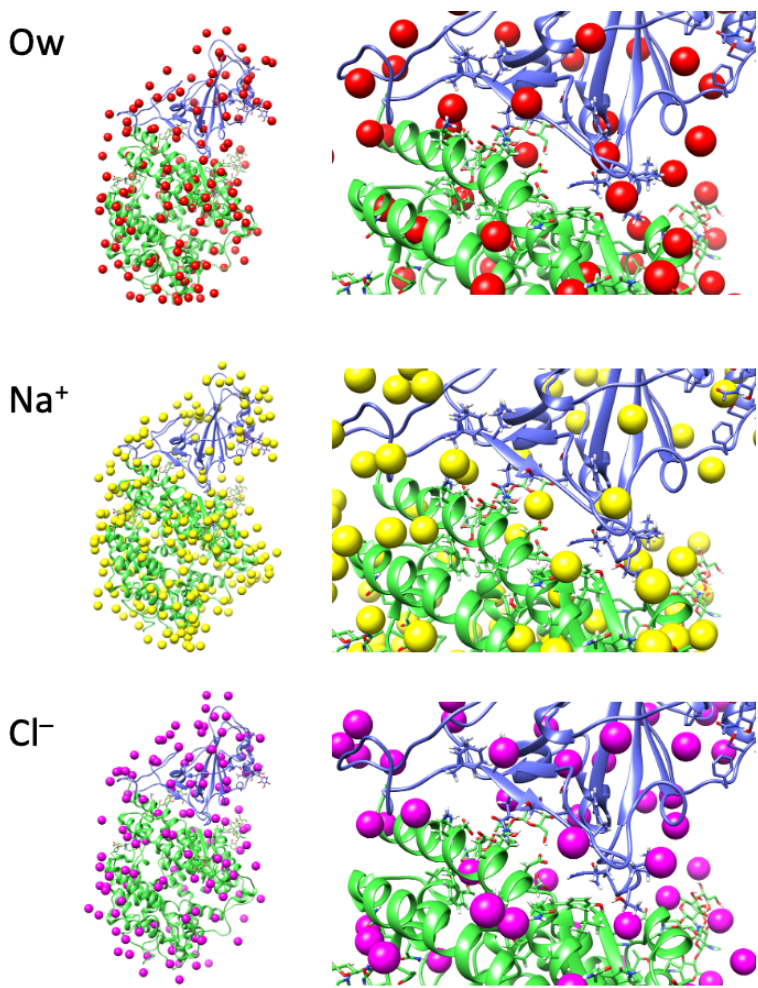

Figure 4. Explicit solvent positions evaluated by the Placevent algorithm. The virtual concentrations of all the solvent species were set to 1.0 M for the Placevent. Oxygen in water and Na$^+$ and Cl$^-$ ions are depicted by red, yellow, and purple spheres, respectively. The panels in the column on the right show a close-up view around the RBD–ACE2 contact region.

**Performance on multiple GPUs**

Before showing the performance of the developed code, the convergence threshold is first discussed. As mentioned above, the RMS residual of the indirect correlation function is employed as a criterion of convergence in RISMiCal, which does not directly indicate the convergence of solvation thermodynamic quantities. The RMS residual of *the j*-th iteration step is defined as

$$\text{RMS}^j = \frac{1}{Nn_v} \sum_i^{\text{Grid}} \sum_\gamma \sqrt{\left(t_\gamma^j(r_i) - t_\gamma^{j-1}(r_i)\right)^2}, \tag{41}$$

where $N$ and $n_v$ are number of grid points and solvent sites, respectively, and the summation is running over all the grid points and solvent sites. The term $t_\gamma^j(r_i)$ is the indirect correlation function of solvent site $\gamma$ of the $j$-th step at grid point $r_i$. Thus, it is better to check the convergence of physical quantities such as SFE or PMV associated with the RMS residual convergence. However, calculating physical quantities at every step is undesirable because it increases the computational cost. In our previous study, a convergence threshold of 1e-6 was set from the SFE of a DNA fragment (754 atoms) in aqueous solution.[95] To check whether this value is affected by the size of the solute or the solvent, the threshold dependences of the SFE and PMV of the RBD–ACE2 complex, ACE2 and RBD in 0.2 M NaCl aqueous solution is shown in Figure 5. The ACE2 (9798 atom) and RBD (3079 atom) structures were used to split the complex one for the calculations. The respective SFEs of the RBD–ACE2 complex, ACE2 and RBD are -13266.65, -11363.23 and -2483.28 kcal/mol when the RMS residual is 1e-9. In Figure 5 (upper), the differences from these values are plotted. The behaviors of the values reaching convergence are different, but they all converge below 1e-6, with differences in SFE at 1e-6 and 1e-9 of < 0.02 kcal/mol. The PMV values for the complex, ACE2 and RBD are 69765.29, 53026.10 and 16656.42 cc/mol, respectively. The differences in PMV at 1e-6 and 1e-9 is also < 0.04 cc/mol, which is not a practical problem. In practice, the convergence behaviors of these physical quantities are system-independent. We have performed numerous 3D-RISM calculations in aqueous ionic solutions with various biomolecules, such as chignolin (138 atom)[111] and its mutants,[38–40,112] Trp-cage (272 atom),[113,114] WW domain (562 atom),[38] cellulose-specific carbohydrate-binding module (572 atom),[115] protein G and its mutant (855 and 846 atom),[40,114] $\alpha$3D (1140 atom),[38] chymotrypsin inhibitor 2 mutant dodecamer (12,684 atom),[97] and AQP1 water channel (15,096 atom).[51,52,54] The threshold value of 1e-6 was sufficient for convergence of the physical quantities in these calculations, so 1e-6 was set as the default value in the codes for single- and multi-GPUs. When performing the 3D-RISM calculations, the calculations at 1e-5, 1e-6, and 1e-7 should be performed to check the calculated physical quantities.

The RMS residual values as a function of the number of iterations are shown in Figure 6. The logarithms of the RMS residual decrease almost linearly with increasing number of iterations: 324 for 1e-6 and 552 for 1e-9 for the complex, 518 for 1e-6 and 763 for 1e-9 for ACE2, and 328 for 1e-6 and 578 for 1e-9 for the RBD. When calculating the 8900 structures in our previous study, the minimum was 299 steps and the maximum 1047, with a threshold value of 1e-6 and a mean value of 385 steps (standard deviation: 70 steps) for the RBD–ACE2 complex. The number of steps at which the RMS residual increases by a factor of 1e-1, e.g., from 1e-5 to 1e-6, depends on the system. The number of steps of convergence also depends on the state and conformation of the structure. In the example calculations, the calculation for the complex converges faster than those for ACE2 and RBD because the complex has a stable structure in aqueous solution, whereas the other two have exposed binding sites. Such calculations that split complexes are performed when estimating the contribution of SFE in the binding energy. However, the linear behavior shown in Figure 6 is the same in all systems. This means that the modified Anderson method has stably converged the 3D-RISM calculations. The $s_1 (= 0.2)$ and $s_2 (= 0.0)$ parameters of the modified Anderson method do not usually need to be changed. If they are roughly in the range $0.2 \leq s_1 + s_2 \leq 0.7$, the convergence behavior does not change significantly. The $b$ parameter might also remain unchanged when using the KH closure, but if it does not converge when using the HNC closure, 0.4 or less can be tried. The MDIIS method also shows similar behavior, as reported previously.[104]

Figure 7 shows the computation time of the above system against the number of GPUs, using the Pegasus supercomputer with one GPU per node. It can be seen that parallel computation scales up to 64 GPUs. The computationally expensive part of the 3D-RISM calculations can be roughly divided into the initialization part, which includes the potential calculation, and the iterative calculation part. As seen in the figure, both parts are efficiently parallelized. When 64 GPUs are used, the total calculation time is 14.6 s (1.4 s for the initialization calculation and 11.4 s for the iteration). The realization of high-speed calculations is a great advantage when performing a large number of the 3D-RISM calculations required for hybrid methods with MD or quantum chemical calculations. Parallel computing also makes more memory available, so the solvation box size can be increased to calculate large protein complexes, and the grid resolution to perform highly accurate solvation thermodynamics.

Figure 8 shows the speedup factors according to the number of GPUs based on the data in Figure 7. Here, the value with two GPUs was used as reference value 1. First, the parallel efficiency of the initialization is good, with 2× acceleration on four GPUs, 3.9× on eight GPUs, and 27.6× on 64 GPUs. This is due to the high independence of the potential calculation at each grid point of the calculation cell with the main contribution. The potential calculation time is proportional to the number of grids and the numbers of solvent components and solute atoms, which is where GPUs excel. Thus, a potential calculation for four solvent components and 12,877 atoms of solute on a $512^3$ grid takes 1.4 s on 64 GPUs; the acceleration rate is higher by a factor of only 27.6 because of the overhead of starting up the GPUs. By contrast, the iteration part involves parallel FFT calculations, which require communications between GPUs. Therefore, even with four GPUs, the acceleration rate is higher by a factor of only 1.3 times compared with two GPUs. The computation time for initialization decreases monotonically as the number of GPUs increases, so the total computation time approaches the iteration time asymptotically. In practice, it is computationally efficient to run with the minimum number of GPUs for which the target system is computationally feasible.

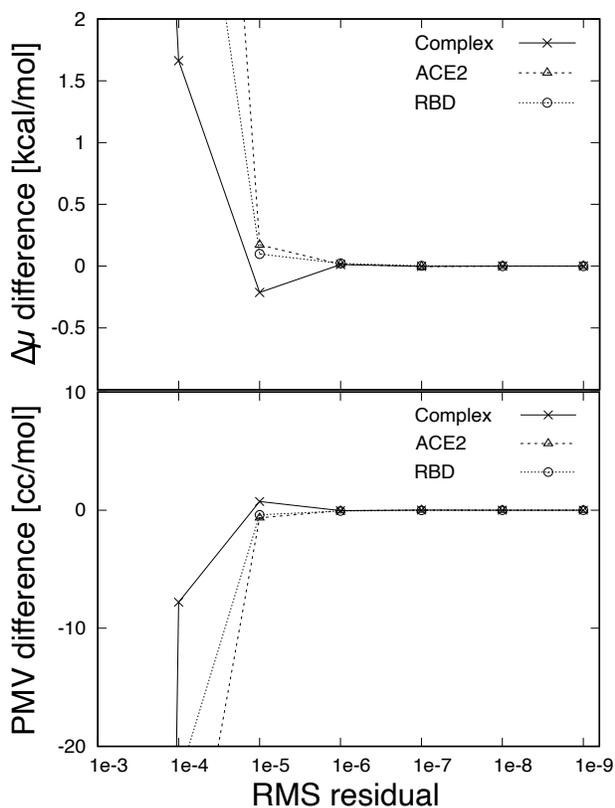

Figure 5. SFE difference (upper) and PMV difference (lower) from the values at threshold 1e-9 of the RBD–ACE2 complex, ACE2, and RBD in 0.2 M NaCl aqueous solvent versus the RMS residual.

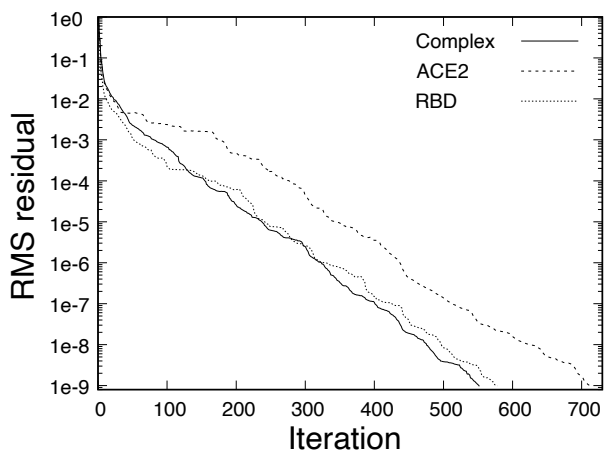

Figure 6. The RMS residual of the RBD–ACE2 complex (solid line), ACE2 (dashed line), and RBD (dotted line) in 0.2 M NaCl aqueous solvent versus the number of iterations.

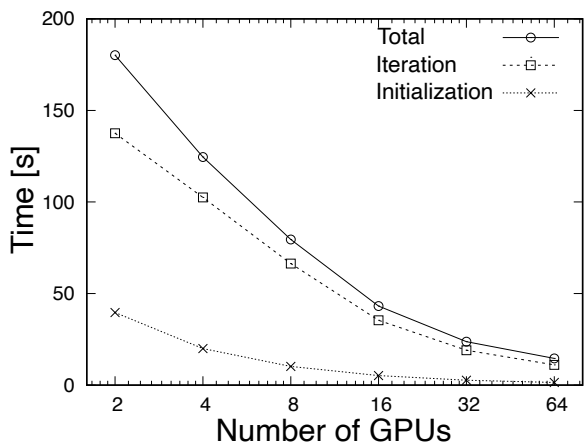

Figure 7. Total calculation, iteration, and initialization times of RBD–ACE2 in 0.2 M NaCl aqueous solvent versus the number of GPUs.

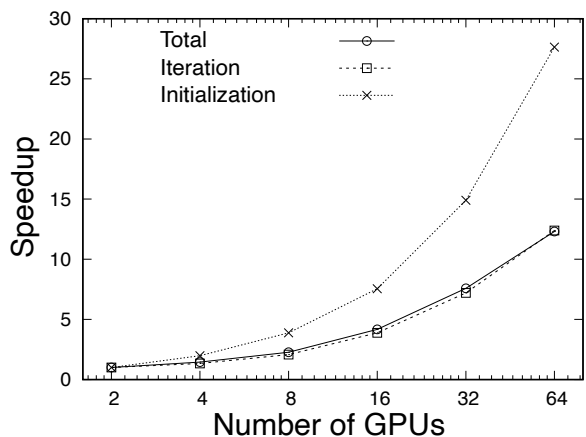

Figure 8. Speedup factors of total calculation, iteration, and initialization parts versus the number of GPUs. The value of two GPUs was used as a reference.

**Summary**

This paper introduces RISMiCal, a program package that enables fast RISM/3D-RISM calculations. The open-source version of RISMiCal implements fast 3D-RISM code for GPGPU-equipped workstations in addition to the 3D-RISM running on standard CPUs. The program can execute parallel jobs for GPGPUs divided into multiple nodes. High parallelization efficiency was also demonstrated in parallel calculations using 64 GPUs. This enables calculations that require large solvent boxes and high-resolution grids.

RISMiCal also provides tools for generating input and analyzing the results. Inputs can easily be generated from Amber, NAMD, and Tinker parameters, as well as PDB coordinate files. Placevent is also available as an analysis tool for the distribution function obtained.

The RISMiCal version to be released in the future will allow the use of RISM-SCF, 3D-RISM-SCF, QM/MM/3D-RISM, and FMO/3D-RISM implemented in GAMESS, which are hybrid methods with quantum chemical calculation methods, as well as MD/3D-RISM in combination with Tinker. In addition to the above, MOZ, solvent polarizable 3D-RISM, and HMC 3D-RISM will be implemented in the future version. There is also a program that runs on massively parallel supercomputers such as supercomputer Fugaku. This version is not open source at this stage, but can be shared in the framework of collaborative research upon request. For the latest information on the RISMiCal package, see https://rismical-dev.github.io.

## ACKNOWLEDGMENTS

This work was financially supported by the Japan Society for the Promotion of Science (JSPS) KAKENHI (Grant Nos. 19H02677, 22H05089, 22H01873, and JP20H03230). Numerical calculations were partially conducted at the Research Center for Computational Science, Institute for Molecular Science, National Institutes of Natural Sciences (Project 22-IMS-C076), and MCRP-S at the Center for Computational Sciences, University of Tsukuba. NY received support from the MEXT Program: Data Creation and Utilization-Type Material Research and Development Project Grant No. JPMXP1122714694. This work was also supported by MEXT as "Program for Promoting Researches on the Supercomputer Fugaku" (Grant Number JPMXP1020230327) and used computational resources of (supercomputer Fugaku provided by the RIKEN Center for Computational Science ) (Project ID: hp230212). Molecular graphics were designed using UCSF Chimera, which was developed by the Resource for Biocomputing, Visualization, and Informatics at the University of California, San Francisco.

## KEYWORDS

reference interaction site model, solvation free energy, graphics processing unit